\documentclass[1p]{elsarticle}

\usepackage{amssymb}

\usepackage{graphicx}
\usepackage{dcolumn}
\usepackage{bm}
\usepackage{upgreek}

\usepackage{epstopdf} 
\begin{document}

\title{Dependence of magnetic domain patterns on plasma-induced differential oxidation of CoPd thin films}

\author[ntnu]{Wei-Hsiang Wang}
\author[ntnu]{Chak-Ming Liu}
\author[nsrrc]{Tzu-Hung Chuang}
\author[nsrrc]{Der-Hsin Wei}
\author[ntnu]{Wen-Chin Lin\corref{cor}}
\ead{wclin@ntnu.edu.tw}
\author[ntnu]{Pei-hsun Jiang\corref{cor}}
\ead{pjiang@ntnu.edu.tw}
\address[ntnu]{Department of Physics, National Taiwan Normal University, Taipei 116, Taiwan}
\address[nsrrc]{National Synchrotron Radiation Research Center, Hsinchu 300, Taiwan}
\cortext[cor]{Corresponding authors}

\begin{abstract}
We demonstrate the evolution of the micro-patterned magnetic domains in CoPd thin films pretreated with e-beam lithography and O$_2$ plasma. During the days-long oxidation, significantly different behaviors of the patterned magnetic domains under magnetization reversal are observed via magneto-optic Kerr effect microscopy on different days. The evolution of the magnetic behaviors indicate critical changes in the local magnetic anisotropy energies due to the Co oxides that evolve into different oxide forms, which are characterized by micro-area X-ray absorption spectroscopy and X-ray photoelectron spectroscopy. 
The coercive field of the area pre-exposed to plasma can decrease to a value 10 Oe smaller than that unexposed to plasma, whereas after a longer duration of oxidation the coercive field can instead become larger in the area pre-exposed to plasma than that unexposed, leading to an opposite magnetic pattern. Various forms of oxidation can therefore provide an additional dimension for magnetic-domain engineering to the current conventional lithographies.
\end{abstract}

\maketitle

\section{Introduction}

Metal-oxide heterostructures have been extensively studied in various aspects, such as interfacial structures and magnetic exchange coupling between ferromagnetic metals and antiferromagnetic oxides \cite{Meiklejohn1956,Stamps2000}, and nanostorage or catalysis applications of nanosized metals on oxide surfaces \cite{Freund2002,Henry1998,Campbell1997}. 
Evolution of metal-oxide growth under different oxidation conditions has also become an important topic \cite{Stierle2007,Williams2013,Galbiati2017,Li2018,Genuzio2020}, including a special focus on self-assembly integrated with conventional photo or e-beam lithography that has been shown to be a powerful tool to achieve complex multiple patterning \cite{Jeong2013}. Since patterned oxidation can be utilized for sophisticated engineering of magnetic structures in thin films \cite{Wang2020}, lithography incorporated with different types of oxide formation is expected to boost the beneficial application of magnetic patterning.

The mechanism of oxidation-related magnetic engineering involves several effects of oxidation on magnetic properties.
Oxygen usually exhibits different affinities with elements of an intermetallic compound, which leads to selective oxidation. This results in self-assembling of the Co-oxide nanoclusters and development of a Pd-rich magnetic layer as observed in our previous study on CoPd alloy thin films \cite{Hsu2017}. With the change in the concentrations of the elements, the magnetic easy direction of the sample gradually turns away from the pristine perpendicular direction to an in-plane direction. Control of the extent of oxidation by varying the thermal annealing time under a partial oxygen pressure, on the other hand, has been found to increase the roughness at the interface of a Co/CoO bilayer structure, leading to domain-wall pinning and an increase of coercivity \cite{Kumar2016}. An enhancement of coercivity is also found possible by implanting oxygen ions into Co films to promote formation of Co oxides at the Co grain boundaries, which reduces exchange interactions between Co crystallites, leading to  a higher coercivity \cite{Demeter2011}.  In contrast to metal/metal oxide bilayers formed by surface oxidation, implantation can result in a different morphology of the interface between the metal and the oxide, and hence may contribute a radically different magnetization mechanism to the system studied \cite{Demeter2011,Menendez2013}. More research is demanded to explore the underlying mechanisms and the full potential of oxidation methodology for magnetic patterning.

In magnetic engineering, ion irradiation has long been considered a reliable process to create magnetic domain patterns with the penetration of keV to MeV ions into the films of magnetic materials \cite{Fassbender2009,Jaafar2011,Bera2020}. Treatments with ions of sufficient energy are believed to be a requirement for successful modification of the magnetic coherence and coupling in a magnetic film in order to generate artificial magnetic microstructures inside the film. However, oxidation-induced magnetic patterning may provide  a more handy method to accomplish magnetic engineering with less invasive treatment.

In this study, we demonstrate oxidation-induced magnetic patterning achieved by surface treatment of CoPd alloy films with low-energy ions generated by a regular plasma cleaning equipment. Evolution of Co oxides on CoPd alloy films pretreated with patterned O$_2$ plasma bombardment is investigated by exploring the behaviors of magnetic domains of the film via magneto-optic Kerr effect (MOKE) microscopy. As oxidation evolves, intriguing different or even opposite behaviors of the patterned magnetic domains are observed upon magnetization reversal, which indicates the existence of various forms of Co oxides during the oxidation, pointing to the possibility of creating complex magnetic structures using conventional lithography that provides directed formation of oxides. Our experiment is the first of its kind to investigate the dependence of the magnetic domains on oxidation forms for exploring the flexibility and diversity of patterning of magnetic properties. 

\section{Material and methods}

The 8-nm-thick CoPd (50:50) alloy film was deposited on SiO$_2$/Si(001) substrates through e-beam-heated coevaporation in an ultrahigh vacuum with a base pressure of $3\times10^{-9}$ mbar. Two evaporation guns were both aligned at 45$^{\circ}$ to the normal. This oblique deposition geometry allows uniaxial magnetic anisotropy to be developed on the surface plane \cite{Chi2012}. The evaporation rates were controlled through the adjustment of the emission currents applied to the sources. 
Alloy composition was determined by the deposition rates of Co and Pd and confirmed through transmission electron microscopy with energy-dispersive spectroscopy (TEM-EDS) \cite{Mudinepalli2017}. A layer of polymethyl methacrylate (PMMA) was then spin-coated on the CoPd film. The PMMA layer was later patterned through e-beam lithography to create a 10-by-10 array of $50 \times 50$ $\upmu$m$^2$ square windows spaced 25 ${\upmu}$m apart. The sample was then treated with 10.5-watt O$_2$ plasma (PDC-32G, Harrick Plasma) for 90 seconds with a fairly low flow rate of 0.1 cc/min of oxygen at a typical base pressure of 200 mTorr of the plasma chamber, followed by removal of PMMA. (According to the Langmuir probe measurement on a Harrick plasma cleaner of a similar model \cite{Nowak1990} and the plasma sheath model \cite{Panagopoulos1999}, the ion energy of the plasma is estimated to be $\sim$50 eV.) After that, the samples were stored in a vacuum desiccator (internal pressure $\sim0.15$ bar) to ensure mild and slow oxidation, which allows us to carefully monitor the different states of magnetism of the sample. The magnetic properties were inspected after fabrication using MOKE microscopy as described in the following contents.

The samples are investigated using longitudinal MOKE microscopy with the external magnetic field ($H$) applied along the magnetic easy axes. All the measurements are carried out at room temperature. The features to be demonstrated in this paper are reproducible between the samples. A lower sweep rate of $H$ with more sampling points per unit field is chosen during the magnetization reversal, where the magnetic domains exhibit faster and more prominent changes. For the MOKE measurements presented in Section \ref{exp},  $H$ is swept at a rate of $\Delta H = \pm10$ Oe or $\pm5$ Oe per 0.5 second for each data point, except for the magnetization reversal processes (respective $|H|$ ranges are 80--120 Oe for Day 0 (Fig.~\ref{0day_m}), 50--100 Oe for Day 10 (Fig.~\ref{10day}), and 0--40 Oe for Day 42 (Fig.~\ref{42day})), during which the sweep rate is decreased to $\Delta H = \pm1$ Oe or  $\pm0.5$ Oe per 0.5 second to gently process the variation of the magnetic domains. The extensively oxidized case on Day 217 (Fig.~\ref{217day}) is an exception, where $H$ is swept at a rate of $\Delta H = \pm5$ per 0.5 second for the whole measurement. All the MOKE images present the same region of the sample, which corresponds to the four squares at the top left corner of the whole array pattern. All the hysteresis loops are measured from the area inside the single square at the top left corner and from the outside area next to it on the right, respectively (see the Supplementary Materials). Note that possible PMMA remnants would not contribute to any field-dependent MOKE signals, as already confirmed in our previous studies \cite{Wang2020}.

\section{Experimental results and discussion}\label{exp}
\subsection{Pinned edge domains (Day 0)}

\begin{figure*}
	\centering
\includegraphics[width=0.85\textwidth]{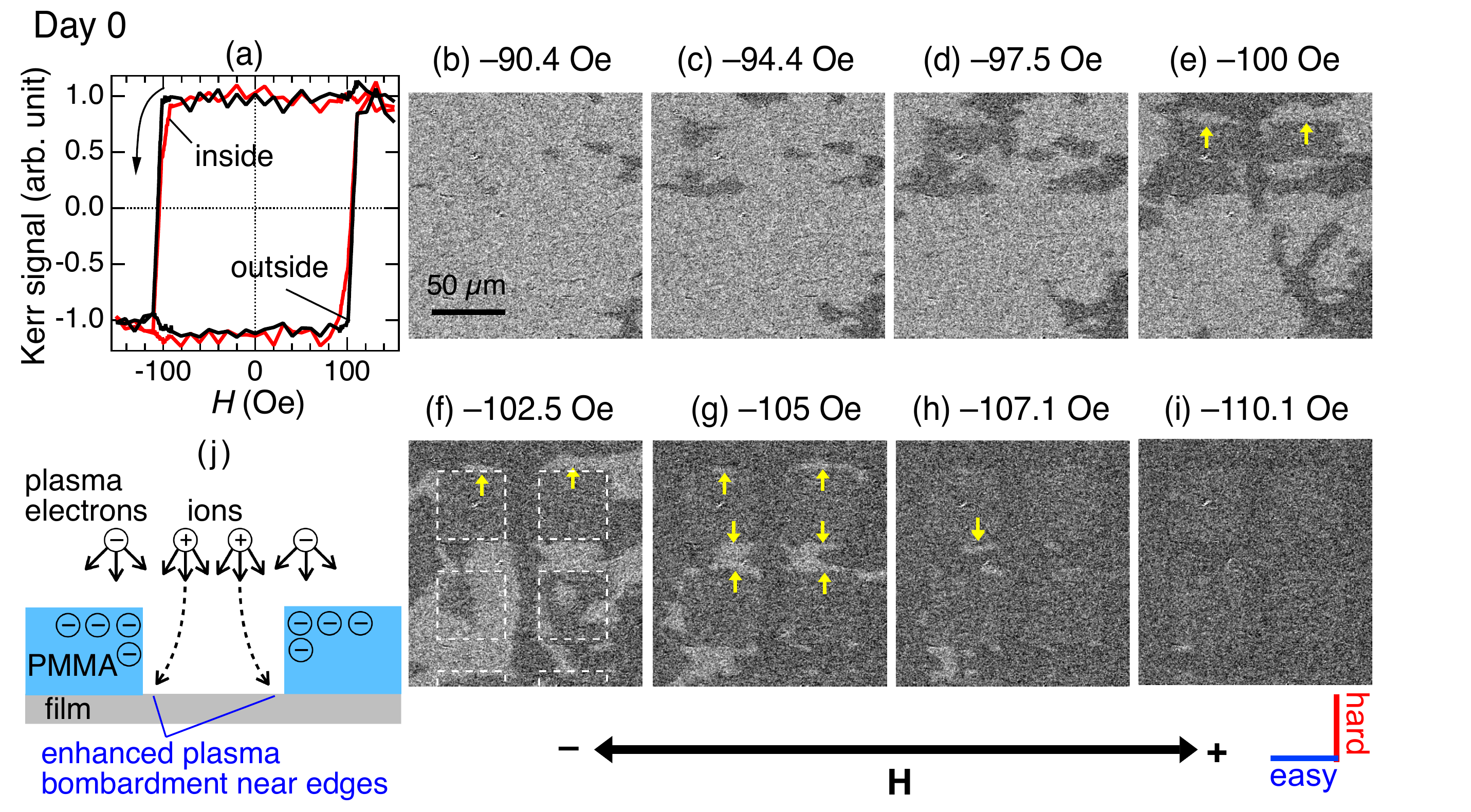}
	\caption{Longitudinal MOKE measurements on Day 0 of the CoPd film pretreated with e-beam lithography and O$_2$ plasma, with the magnetic field $H$ applied along the easy axis.  (a) The hysteresis loops of the areas inside and outside a plasma-pretreated $50 \times 50$ $\upmu$m$^2$ square, respectively. (b)--(i) Differential MOKE images of four squares showing the evolution of the magnetic domains as $H$ is varied. Square edges are marked by white dashed lines in (f). (j) Illustration of deflected ions during plasma treatment due to the negatively charged PMMA (not to scale).}\label{0day_m} 
\end{figure*}

Presented in Fig.~\ref{0day_m} is the representative sample measured right after the e-beam lithography and O$_2$ plasma treatment. Fig.~\ref{0day_m}(a) shows the hysteresis loops of the areas inside (red curve) and outside (black curve) a square, respectively, and Figs.~\ref{0day_m}(b)--(i) show the MOKE images of four squares among the entire square array on the film simultaneously recorded with Fig.~\ref{0day_m}(a) as $H$ is changed gradually from $-90.4$ Oe to $-110.1$ Oe. 
In Fig.~\ref{0day_m}(a), no appreciable difference is found between the curves inside and outside a square, which is consistent with the observation that the square array pattern cannot be clearly recognized in Figs.~\ref{0day_m}(b)--(e), while the total size of the dark domains with magnetization pointing to the negative direction gradually increases as $H$ is changed from $-90.4$ to $-100$ Oe. However, for $H = -100$ to $-107.1$ Oe, as shown in Figs.~\ref{0day_m}(e)--(h), hints of persisting light-gray domains are observed along the edges of the squares, which are indicated with yellow arrows. These line segments of \textit{pinned}, light domains persist until a more negative $H$ is applied, which eventually results in a completely dark image shown in Fig.~\ref{0day_m}(i) at $H=-110.1$ Oe. The magnetic behavior inside a square does not differ much from that outside, which indicates that the magnetic properties of the sample do not change significantly right after the plasma treatment, except that the properties near the edges of the squares seem to be altered, leading to the pinned magnetic domains near the edges. The light-gray domain segments are more prominent along the edges parallel to the easy axis of the film than that along the hard axis.

This phenomenon is reminiscent of the edge domain traps observed in an experiment of Bauer \textit{et al.~}\cite{Bauer2013} with voltage-controlled oxidation states in their device, in which oxidation of the Co layer is most efficient at the \textit{edge} of the GdO$_x$ gate dielectric, where the triple phases (i.e., O$_2$ gas, O$^{2-}$ ion-conducting, and electron-conducting phases) meet \cite{Mogensen1996, Balluffi2005}. In our experiment, the pinned edge domains are probably also caused by the reactions being most efficient at the edges, possibly due to trajectories of positive ions being deflected towards the negatively charged PMMA sidewalls \cite{Watanabe2001,Ishchuk2012}, leading to enhanced bombardment of O$_2$ plasma at the square edges, as illustrated in Fig.~\ref{0day_m}(j). Denser defects are thereby formed along the edges. It has been shown in previous studies that small-scale defects in magnetic materials can act as pinning centers for domain wall movement \cite{Paul1982,OHandley2000,Burn2014,Demeter2011}. This explains the domain wall pinning observed along the square edges.

A vague contrast between the areas inside and outside the squares can be observed in Fig.~\ref{0day_m}(i) when the sample is magnetically saturated. (This slight contrast can be more clearly seen later in Fig.~\ref{10day}(i) on Day 10.) The remnant contrast under saturation is often detected in our studies on CoPd or FePd thin films with their magnetic domain patterns created using lithography and plasma treatments. Further investigation of these samples using zero-field polar MOKE microscopy shows clear contrasts of the squares \cite{Wang2020}, which reveal the difference in the perpendicular magnetization between and areas inside and outside the squares. This may explain the existence of the magnetic contrast between the two regions even under saturation.

\subsection{Preceding magnetization reversal in plasma-pretreated areas (Day 10)}

\begin{figure*}
	\centering
\includegraphics[width=0.85\textwidth]{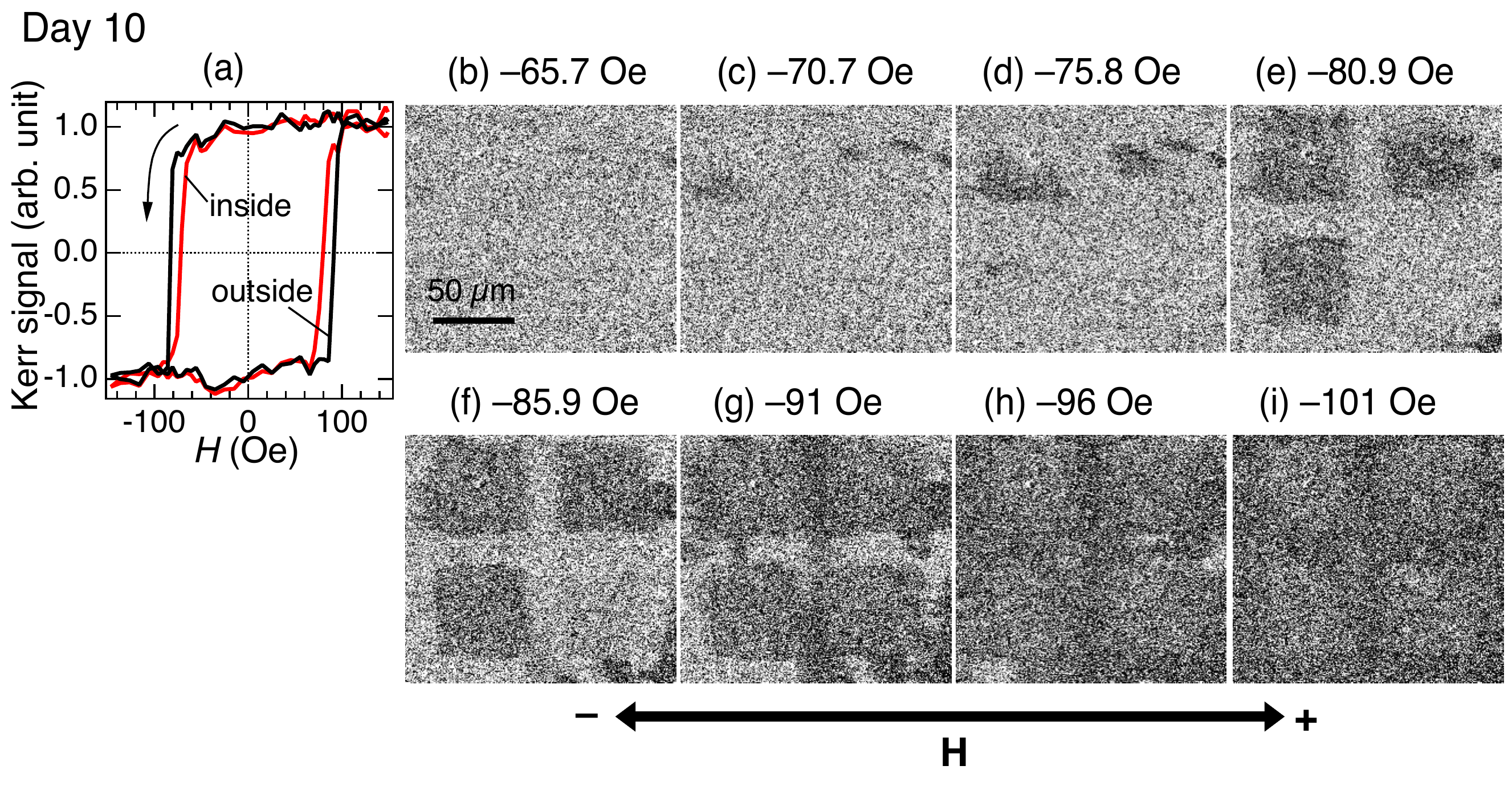}
	\caption{Longitudinal MOKE measurements of the CoPd film on Day 10. (a) The hysteresis loops of the areas inside and outside a square, respectively. (b)--(i) Differential MOKE images of four squares showing the evolution of the magnetic domains as $H$ is varied.}\label{10day} 
\end{figure*}

The sample is re-measured after 10 days of storage in a vacuum desiccator after the O$_2$ plasma treatment. The hysteresis loops for the areas inside (red) and outside (black) the top left square are shown in Fig.~\ref{10day}(a). The coercivity ($H_\mathrm{c}$) decreases from 105 Oe as shown in Fig.~\ref{0day_m}(a) to 85 Oe for outside and to 75 Oe for inside a square, respectively, after mild oxidation for 10 days. (Statistics of the $H_\mathrm{c}$ values over the whole array pattern and video recordings of the MOKE images are provided in the Supplementary Materials.) Local magnetic properties are also modified by oxidation, as revealed in the magnetic patterns observed in the MOKE images in Figs.~\ref{10day}(b)--(i). During the reversal of the magnetization of the film from the positive (light gray) to the negative (dark gray) direction as $H$ is changed from $-70.7$ Oe to $-96$ Oe, the magnetization inside the plasma-pretreated squares is reversed before that outside the squares, exhibiting dark square domains under a mediate $H$ around $-85.9$ Oe. The observed magnetic pattern can be explained by a higher rate of surface oxidation induced by plasma pretreatment. Ion bombardment is known to stimulate production of point defects, such as vacancies and interstitials \cite{Williams1984}, which enhance the mobility and reactivity of oxygen anions and metal cations in materials. Areas inside the squares thereby have a higher surface oxidation rate, which leads to a lower magnetic anisotropy energy (MAE) and hence a smaller $H_\mathrm{c}$  \cite{Gambardella2004,Zhang2011a,Hsu2017,Girt2006,Zhang2013a} inside the squares than that outside, as shown in Fig.~\ref{10day}(a). Therefore, magnetization reversal is easier inside the squares than outside. 

\subsection{Delayed magnetization reversal in plasma-pretreated areas (Day 42)}

\begin{figure*}
	\centering
\includegraphics[width=0.85\textwidth]{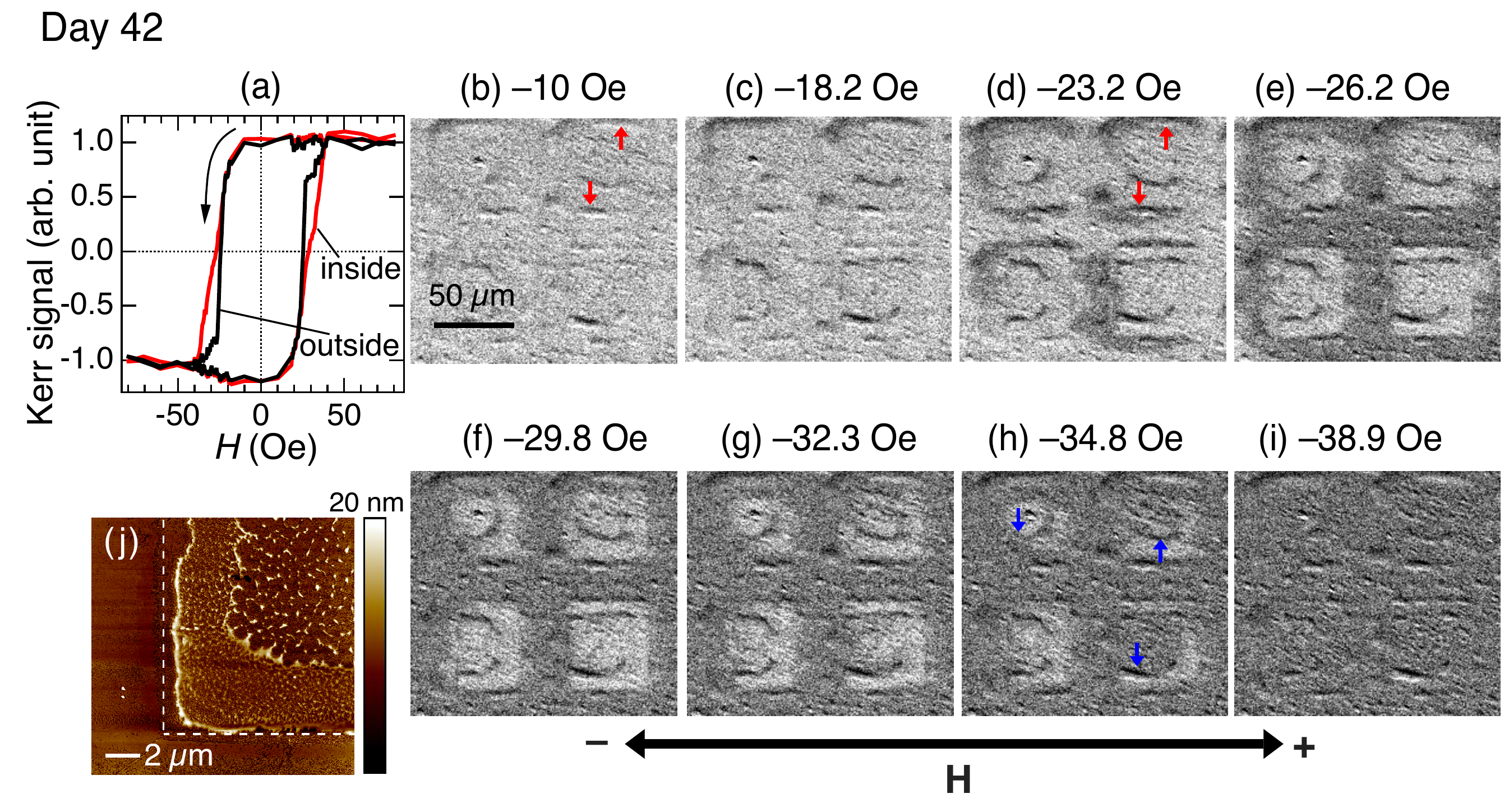}
	\caption{Longitudinal MOKE measurements of the CoPd film on Day 42. (a) The hysteresis loops of the areas inside and outside a square, respectively. (b)--(i) Differential MOKE images of four squares showing the evolution of the magnetic domains as $H$ is varied. (j) Atomic force microscopy image of a corner of a square, with the square edges marked by dashed white lines. A ridge of the Co oxide clusters can be clearly seen along the edges inside the square.}\label{42day}
\end{figure*}

The sample is measured again after 42 days of storage in a vacuum desiccator after the O$_2$ plasma treatment. An intriguing \textit{opposite} effect of oxidation on $H_\mathrm{c}$ is observed. 42-day oxidation reduces $H_\mathrm{c}$ outside the squares \textit{more} than it does inside the squares, as presented in Fig.~\ref{42day}(a) with $H_\mathrm{c}$ inside the top left square (29 Oe) larger than $H_\mathrm{c}$ outside (25 Oe), which is contrary to the result of 10-day oxidation shown in Fig.~\ref{10day}(a). (Comparison of the dynamic behaviors of the magnetic domains displayed in Figs.~\ref{10day} and \ref{42day} is more clear in video recordings provided in Videos S1 and S2 in the Supplementary Materials.) To investigate the cause of this opposite effect, the MOKE images (Figs.~\ref{42day}(b)--(i)) are examined closely as follows. The red arrows in Figs.~\ref{42day}(b) and \ref{42day}(d) indicate some of the segments of local oxide clusters grown near the edges of the squares, which can be seen more clearly in an atomic force microscopy image shown in Fig.~\ref{42day}(j). Also observed in Fig.~\ref{42day}(j) are small nanodots of oxides away from the edges that have not self-assembled to form larger clusters \cite{Hsu2017}. When $H$ is changed to more negative values, dark domains are first established near the edge oxide clusters (see Fig.~\ref{42day}(d) at $-$$23.2$ Oe), demonstrating that these local oxide clusters behave like domain-wall traps. As opposed to the Day-10 case where the MAE is uniformly reduced inside the squares, 42-day oxidation specifically promotes the growth of oxide clusters near the edges, indicating intensive local oxidation there, which further reduces the MAE along the edges. Unlike a minor reduction of MAE along the edges that only pins the magnetic domains on the edges as shown in Fig.~\ref{0day_m} when the sample is measured immediately after the plasma treatment, a local \textit{deep} well of MAE shaped in a closed loop can result in an enhancement in $H_\mathrm{c}$ for the entire enclosed area due to the strong trapping effect of the MAE well on the domain walls \cite{Bauer2015}. Therefore, 
the magnetic domain walls along all the edges of the squares are pinned as the magnitude of $H$ increases, which leads to a delayed reversal of the magnetization inside the squares as shown in the MOKE images in Figs.~\ref{42day}(b)--(i). The tendency for the growth of oxide clusters to start near the edges is likely due to the enhanced plasma bombardment mentioned in Fig.~\ref{0day_m}(j). There are other oxide clusters randomly located inside the squares. These spots also behave like domain-wall traps, as indicated in Fig.~\ref{42day}(h) with blue arrows. 

This experimental result is somewhat different from our previous study on a naturally oxidized CoPd film without plasma pretreatment \cite{Hsu2017}, where no relation between the magnetic domain walls and the locations of the Co-oxide clusters was observed in that film via MOKE microscopy measurements. In a naturally oxidized CoPd film, 
a uniform Pd-rich layer was formed in the CoPd film because of diffusion of the Co atoms to the surface to create a Co-oxide layer with some self-assembled Co-oxide clusters distributed on it. TEM-EDS elementary mapping confirmed that the concentrations of respective elements of the naturally oxidized alloy film stayed quite uniform over the whole film despite the presence of local Co-oxide clusters that were self-assembled on the oxide surface. This resulted in uniform magnetic properties of the film with no domain walls observed at the locations of the Co-oxide clusters. This is different from our present results with plasma pretreatment, where domain walls are established along the Co-oxide clusters inside the squares (Figs.~\ref{42day}(b)--(i)). This discrimination implies an essential difference in the properties of the Co-oxide clusters between the CoPd films with plasma pretreatment and those without it. 

\subsubsection*{Surface characterization}

\begin{figure}
	\centering
\includegraphics[width=0.6\textwidth]{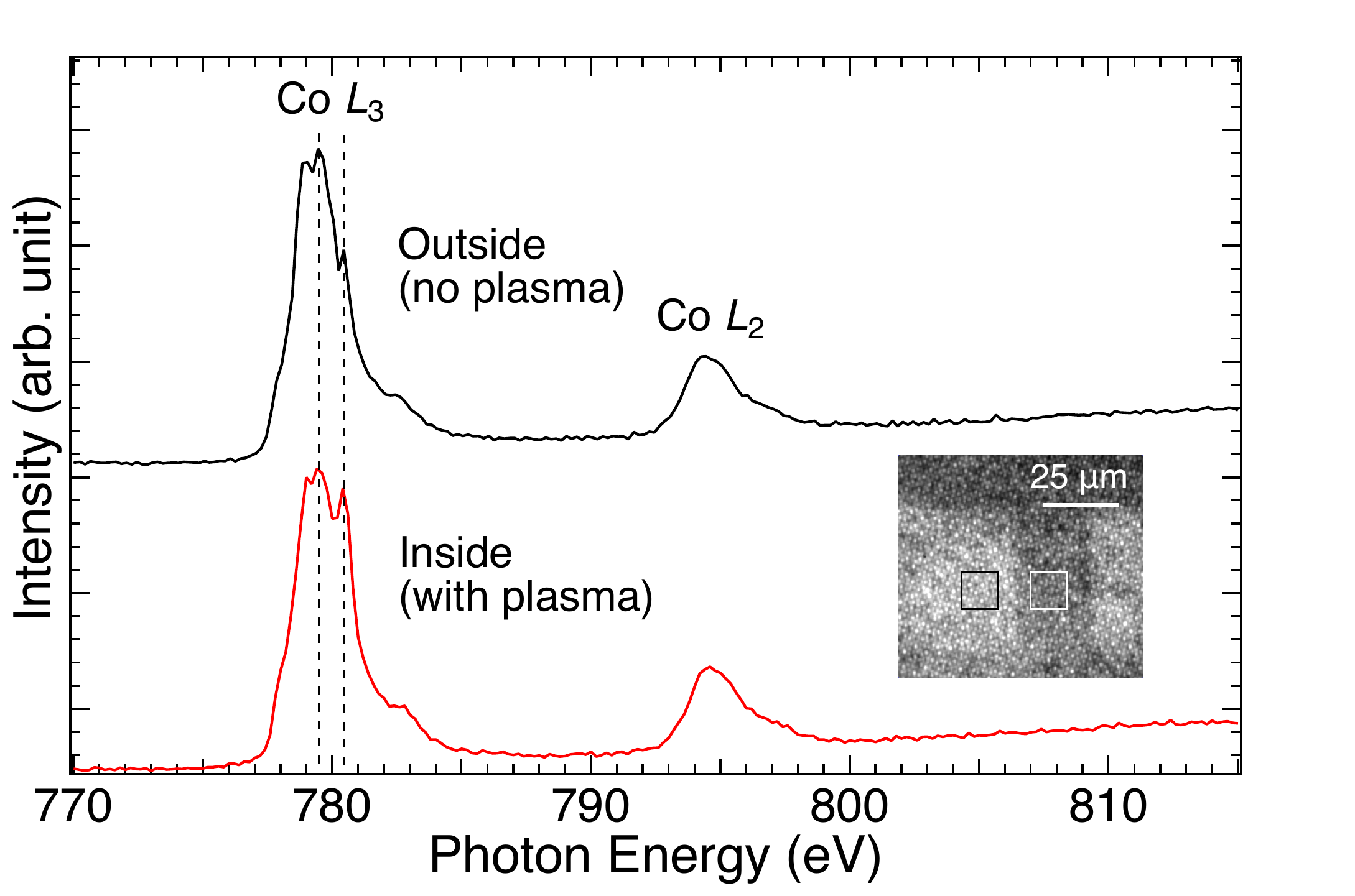}
	\caption{$\upmu$-XAS curves of the film measured from inside and outside a square, respectively. Curves are offset vertically for clarity. The Co \textit{L}$_3$ peaks at 779.5 and 780.5 eV indicate a mixture of Co and CoO, and a higher intensity of the 780.5-eV peak inside a square than outside implies a more extensive oxidation inside. The inset is an XMCD image of the film at 779 eV from which the $\upmu$-XAS data are extracted.}\label{XAS} 
\end{figure}

In order to evaluate the Co oxides at this stage, micro-area X-ray absorption spectroscopy ($\upmu$-XAS) and X-ray photoelectron spectroscopy (XPS) are used to inspect the areas inside and outside the squares of a CoPd sample with a state of oxidation and magnetic property equivalent to that shown in Fig.~\ref{42day} (Day 42). Fig.~\ref{XAS} is the $\upmu$-XAS results extracted from X-ray magnetic circular dichroism photoemission electron microscopy (XMCD-PEEM) at beamline BL05B2 of the Taiwan Light Source in Hsinchu, Taiwan. The top black curve is obtained from an area outside a square, and the bottom red curve is from inside a square, with the source areas indicated in the inset XMCD image by the boxed regions. Both curves show partial oxidation of the metallic Co to CoO, as evidenced by a broad peak around 780 eV that comprises two Co \textit{L}$_3$ peaks at 779.5 and 780.5 eV, which indicate a mixture of Co and CoO \cite{Tan2019}. The major difference between the two curves is that the relative intensity of the 780.5-eV peak inside the square is higher than outside. This indicates more extensive oxidation of the areas exposed to plasma pretreatment than those protected from plasma by PMMA masking. The result is consistent with our optical and atomic force microscopic observations of the 
cluster structures grown inside the squares after mild oxidation during the sample storage. Oxidation thereby is considered a key mechanism that leads to different magnetic properties inside and outside the squares.

\begin{figure}[bt!]
	\centering
\includegraphics[width=0.68\textwidth]{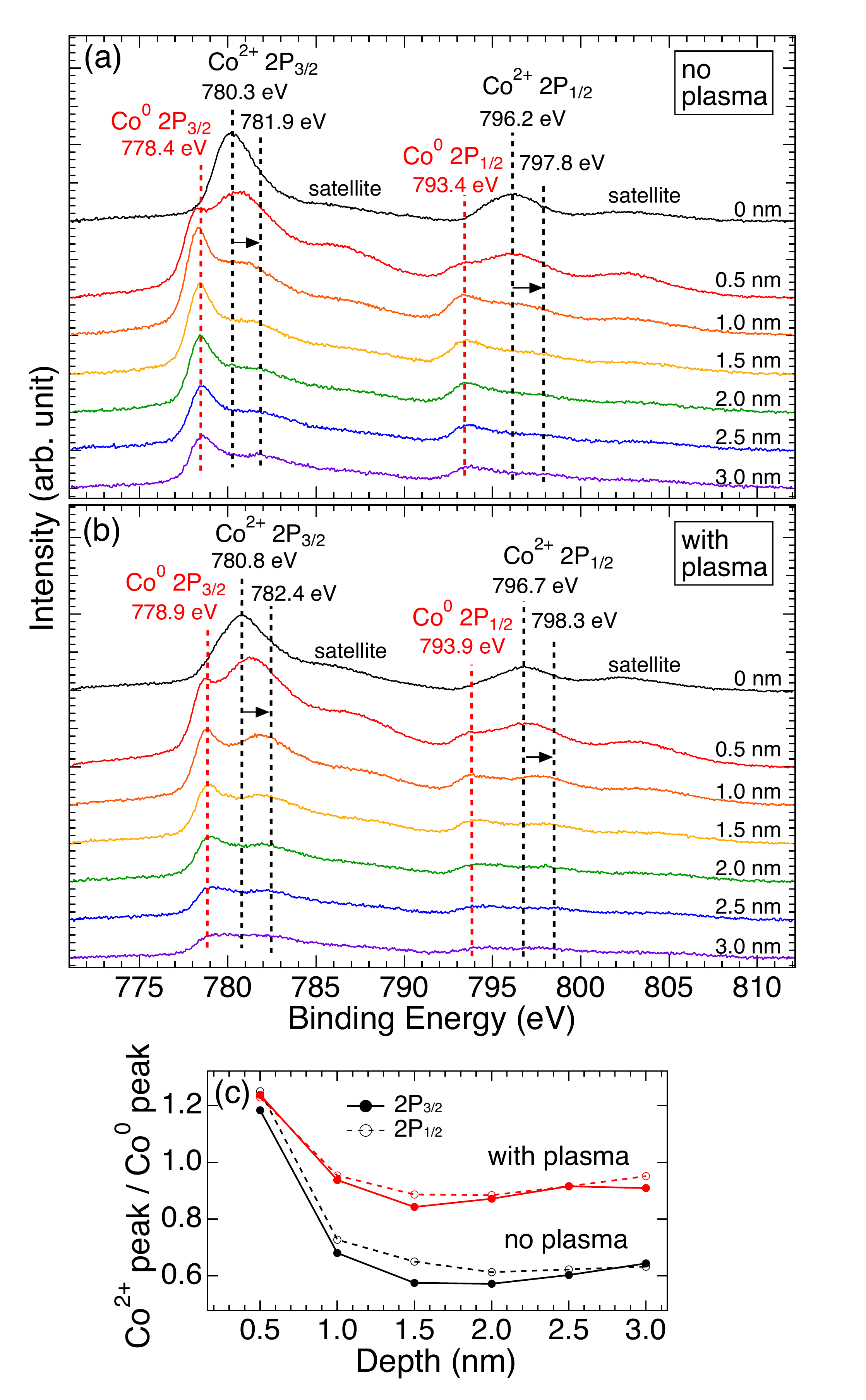}
	\caption{XPS depth profile of the film (a) without and (b) with plasma pretreatment, respectively. Curves from different depths are offset vertically for clarity. The spectra from deeper (1--3 nm) locations show a higher relative intensity of the Co$^0$ peaks with respect to the Co$^{2+}$ peaks in Fig.~\ref{XPS}(a) than in Fig.~\ref{XPS}(b), with the peak ratios shown in (c). This indicates that the oxidation penetration is deeper in the plasma-pretreated areas than in those without plasma treatment.}\label{XPS} 
\end{figure}

Figs.~\ref{XPS}(a) and \ref{XPS}(b), on the other hand, display the depth-dependent XPS spectra from the areas without and with plasma pretreatment, respectively, with Co$^0$ (from the Co metal) and Co$^{2+}$ (from CoO) peaks detected in the measurements. Each figure has seven curves representing the XPS data at 0 to 3 nm in depth from the film surface, with a sputtering depth resolution of 0.5 nm. It can be seen from the deeper (1--3 nm) curves that the relative intensity of the Co$^{2+}$ peaks with respect to the Co$^0$ peaks are higher in Fig.~\ref{XPS}(a) than in Fig.~\ref{XPS}(b), with the peak ratios of Co$^{2+}$ to Co$^0$ shown in Fig.~\ref{XPS}(c). This indicates that the oxidation penetration is deeper in the areas with plasma pretreatment than in those without it. Another important message from the XPS data is the energy shifts of the peaks. There exist two kinds of major energy shifts in the system. First, all the Co$^0$ and Co$^{2+}$ peaks shift to a higher energy with plasma pretreatment with $\Delta E$ $\sim 0.5$ eV. Secondly, the Co$^{2+}$ peaks at 3 nm are $\sim$1.6 eV higher than those at the surface (0 nm). The shift in binding energies with plasma pretreatment indicates changes in the electronic structures of the Co metal and oxide induced by plasma pretreatment \cite{Yu2018}, which may depict the structural difference in the oxide forms between the CoPd alloy films without and with plasma pretreatment. The higher binding energy of the internal Co valance state than that of the surface state, on the other hand, may be attributed to the lower coverage of the Co oxides inside the film than that on the surface \cite{Diaz-Fernandez2014}. As oxidation proceeds, Co oxides tend to evolve from CoO into Co$_3$O$_4$ and Co$_2$O$_3$, and the morphology evolves from scattered oxide clusters into a whole oxide layer. The Co$^{2+}$ peak can be mixed with a few Co$^{3+}$ signals, which have a lower binding energy than Co$^{2+}$. The corresponding binding energy of the combined peak thereby shifts to a lower value as the portion of Co$^{3+}$ increases when moving from deep inside the film to the surface.

To interpret the XPS data, an important consideration is whether \textit{internal} oxidation \cite{Stott1988} or \textit{surface} oxidation of Co occurs in the samples. Surface oxidation is known to yield an oxide layer on the surface that provides protection to some extent \cite{Wood1987}. For CoPd films without plasma pretreatment, Co atoms diffuse to the alloy surface and react with O$_2$ in the air to develop a complete Co-oxide surface layer, on which self assembly occurs later to build Co-oxide clusters. For CoPd films with O$_2$ plasma pretreatment, however, diffusion of oxygen ions into the film during the plasma treatment promotes internal oxidation from the start before a complete surface oxide can be formed on the surface to provide protection. This explains the deeper penetration of the oxidation and the overall change of the electronic structures observed from the XPS. Our MOKE observations in Fig.~\ref{42day} are closely related to the nonuniform distribution of the oxide in the film due to internal oxidation. It has been known that an antiferromagnet (AFM) can enhance the $H_\mathrm{c}$ of an adjacent ferromagnet (FM) even above the Néel temperature ($T_\mathrm{N}$) of the AFM \cite{Leighton2002}. This phenomenon is attributed to an additional anisotropy in the FM induced by the spin fluctuations in the AFM above its $T_\mathrm{N}$. The embedded CoO induced by the implantation of oxygen ions in our experiment is known to be an AFM with a $T_\mathrm{N}$ of 298 K, surrounded by the CoPd film, which is an FM \cite{Leighton2002,Demeter2010,Demeter2011}. Therefore, the enhancement of $H_\mathrm{c}$ inside the squares pretreated by O$_2$ plasma can also be attributed to the formation of internal Co oxides in the CoPd film inside the squares, which act as pinning centers for the domain walls, giving a higher $H_\mathrm{c}$ to the neighboring area around the oxide clusters \cite{Demeter2010}.

\subsection{Extensive oxidation (Day 217)}

\begin{figure}
	\centering
\includegraphics[width=0.65\textwidth]{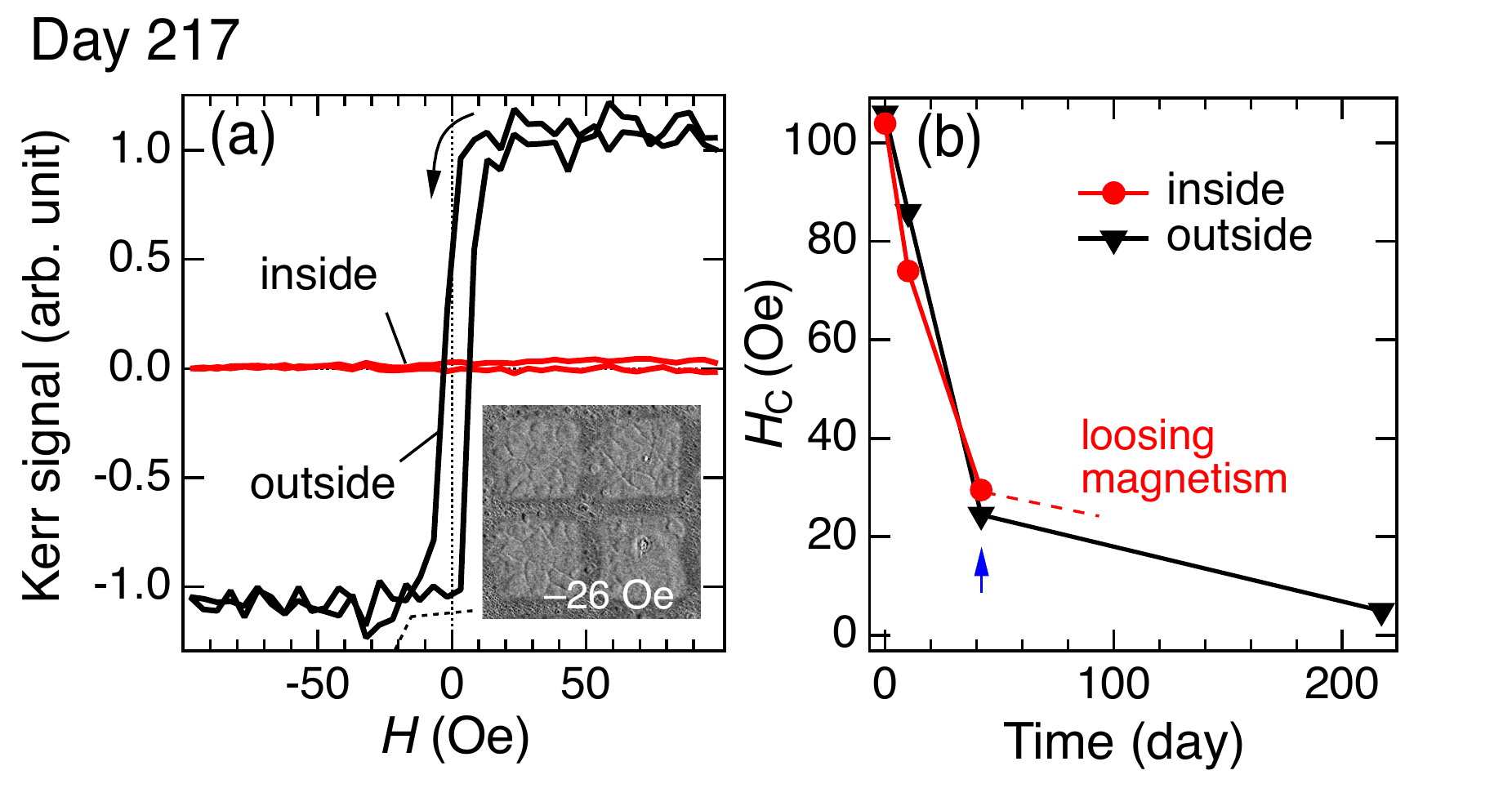}
	\caption{(a) The hysteresis loops of the areas inside and outside a square, respectively, on the CoPd film on Day 217. The inset shows an example of the differential MOKE image at $-26$ Oe. (b) Coercivity $H_\mathrm{c}$ as a function of time for areas inside and outside a square, respectively. $H_\mathrm{c}$ generally stays lower inside a square than outside, except that on Day 42 as indicated by an arrow, $H_\mathrm{c}$ becomes higher inside a square.}\label{217day} 
\end{figure}

The sample is measured again after 217 days of storage in a vacuum desiccator after the O$_2$ plasma treatment. The hysteresis loops inside and outside the top left square are shown in Fig.~\ref{217day}(a). It can be seen that the $H_\mathrm{c}$ of the area outside the squares decreases further to 5 Oe. For the area inside a square, on the other hand, no magnetization is observed at all. This implies that the interiors are almost completely oxidized due to internal oxidation or intergranular penetration \cite{Wood1987} that have been actively taking place during this long time. Deeper penetrations of the internal oxidation in the alloy can occur near the grain boundaries, which promotes the degradation of the sample.

The dependence of $H_\mathrm{c}$ on the duration of oxidation is summarized in Fig.~\ref{217day}(b). As $H_\mathrm{c}$ decreases with time, one can see that $H_\mathrm{c}$ inside the squares generally stays lower than that outside, except for the Day-42 case, where $H_\mathrm{c}$ inside is higher, exhibiting an opposite effect of evolving oxidation on $H_\mathrm{c}$, which leads to a reversed outcome of the domain patterns on Day 42. The difference between $H_\mathrm{c}$ inside the square (exposed to plasma) and that outside (without plasma exposure) is largest on Day 10 with $\Delta H_\mathrm{c}$ = 10 Oe, which has a magnitude comparable with the decrement in $H_\mathrm{c}$ accomplished by 2-keV ion beam erosion \cite{Bera2020}. It is quite intriguing that modification of the surface MAE with the surface plasma treatment can determine the magnetic domains of the whole continuous magnetic film. As the duration of oxidation increases from 0 to 10 days and then to 42 days, the oxidation mechanism that is most responsible for the featured magnetic properties on respective days evolves from edge, to surface, and then to internal oxidation. This demonstrates the potential of plasma-triggered oxidation for enhancing the variety of magnetic patterning with variable oxidation breadths and depths. If further incorporated with precisely controllable or even reversible oxidation techniques such as a gating structure that enables voltage-induced oxidation and reduction \cite{Bauer2013,Zehner2020}, oxidation-induced domain patterning can be a powerful technique for future data-storage and spintronic devices.

\section{Conclusions}

In summary, we demonstrate the oxide-dependent evolution of the magnetic domain patterns under magnetization reversal of a CoPd thin film that is pretreated with e-beam lithography and O$_2$ plasma. MOKE measurements reveal different effects of oxidation on the magnetic micro-domains at different stages of the Co-oxide evolution. During the days-long oxidation process in a vacuum storage desiccator, the dynamic behavior of the magnetic micro-patterns under magnetization reversal evolves into different phases. In the beginning (Day 0), pinned segments of magnetic domains are observed along the edges of the plasma-pretreated $50 \times 50$ $\upmu$m$^2$ squares.  As oxidation continues, the film then exhibits preceding reversal of the magnetization inside the squares on Day 10, but later demonstrates an opposite behavior with delayed magnetization reversal inside the squares on Day 42.  On Day 217, the magnetism inside the squares totally disappears after extensive oxidation. The mechanism behind these phenomena involves different forms of surface and internal oxidation of Co in the film, leading to different alterations of the local magnetic anisotropy energy and coercivity. Various forms of oxidation provide an additional dimension of magnetic patterning to the existing conventional lithography. Magnetic-domain engineering via plasma-induced differential oxidation may be applied to design and fabricate future data-storage and spintronic devices.

\section*{Acknowledgment}

We acknowledge the group of Prof. Hsiang-Chih Chiu for assisting us with the plasma source. This study is sponsored by the Ministry of Science and Technology of Taiwan under Grants Nos.~MOST 107-2112-M-003-004, MOST 108-2112-M-003-011-MY2, MOST 109-2112-M-003-009, and MOST 110-2112-M-003-019.

\bibliography{CoPd}
\bibliographystyle{elsarticle-num-names}
\newpage

\end{document}